# Spatial coherence of electron beams from field emitters and its effect on the resolution of imaged objects

Tatiana Latychevskaia

Physics Department of the University of Zurich

Winterthurerstrasse 190, 8057 Zurich, Switzerland

*Corresponding author: tatiana@physik.uzh.ch

**Abstract**

Sub-nanometer and nanometer-sized tips provide high coherence electron sources. Conventionally, the effective source size is estimated from the extent of the experimental biprism interference pattern created on the detector by applying the van Cittert Zernike theorem. Previously reported experimental intensity distributions on the detector exhibit Gaussian distribution and our simulations show that this is an indication that such electron sources must be at least partially coherent. This, in turn means that strictly speaking the Van Cittert Zernike theorem cannot be applied, since it assumes an incoherent source. The approach of applying the van Cittert Zernike theorem is examined in more detail by performing simulations of interference patterns for the electron sources of different size and different coherence length, evaluating the effective source size from the extent of the simulated interference pattern and comparing the obtained result with the pre-defined value. The intensity distribution of the source is assumed to be Gaussian distributed, as it is observed in experiments. The visibility or the contrast in the simulated holograms is found to be always less than 1 which agrees well with previously reported experimental results and thus can be explained solely by the Gaussian intensity distribution of the source. The effective source size estimated from the extent of the interference pattern turns out to be of about 2-3 times larger than the pre-defined size, but it is approximately equal to the intrinsic resolution of the imaging system. A simple formula for estimating the intrinsic resolution, which could be useful when employing nano-tips in in-line Gabor holography or point-projection microscopy, is provided.

## 1. Introduction

In coherent imaging with electrons, the electron wave is provided by the field emission from a very sharp tip, which ensures high coherence. The temporal coherence is determined by the electron energy spread $l_{\text{coh}}^{\text{Temporal}} = \lambda^2 / \Delta\lambda = 2\lambda U / \Delta U$, where $(eU \pm e\Delta U)$ is the energy of electrons, which is usually relatively high. For example, for $eU = 250$ eV and an energy $e\Delta U = 0.1$ eV spread, the temporal coherence is $l_{\text{coh}}^{\text{Temporal}} = 390$ nm. Thus, typical temporal coherence length exceeds typical sizes of objects and is therefore sufficient for coherent imaging. The study presented here therefore considers monochromatic sources. The spatial coherence of the electron wave is determined by the effective source size, and therefore is always finite. The subject of this study is the "somewhat obscure question of coherence length," as Gabor called it [1]. Some general discussions on spatial coherence can be found in the literature [2-4].

The effective size of the source that provides a coherent wave is conventionally estimated from the extent of the interference pattern in an experimental image, and by applying the van Cittert-Zernike theorem, as demonstrated in [5-8]; we will hereafter refer to this approach as Method 1. The estimated effective source size is typically smaller than the physical size of the tip. In this study, we simulate the electron wave emitted from a coherent source, an incoherent source and a partially coherent source. We discuss the applicability of the van Cittert-Zernike theorem; finally, we study the effect of the source size and source vibrations on the resolution of the imaging system.

## 2. Intensity distribution from a Gaussian-distributed intensity source

The intensity distribution of an electron beam extracted by field emission from a nano tip [9-10] in the detector plane follows a Gaussian distribution, as has been observed experimentally [7, 11-12], which can be described by:

$$I_D(X,Y) = I_{D,0} \exp\left(-\frac{X^2 + Y^2}{2\sigma_D^2}\right), \tag{1}$$

where $\sigma_D$ is the standard deviation, $(X,Y)$ are the coordinates in the detector plane and $I_{D,0}$ is the maximum of the intensity in the centre of the detector. The arrangement of the experiment and the labelling of the planes and coordinates are illustrated in Fig. 1.

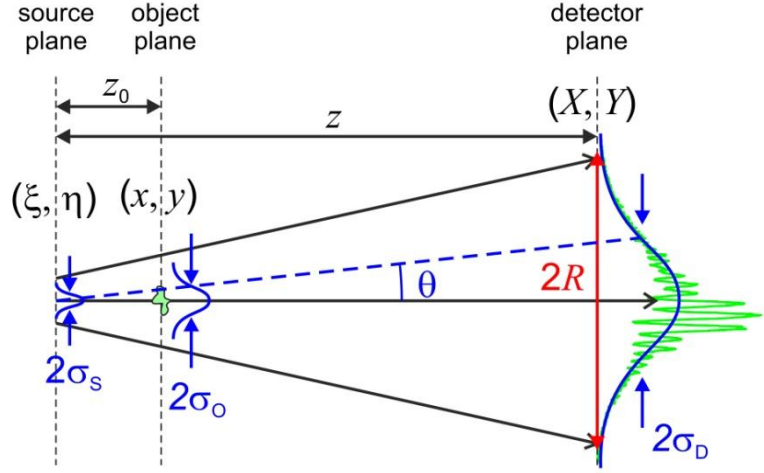

Fig. 1. Illustration of intensity distribution in different planes and coordinate notation.

In order to verify how the coherence of the source affects the appearance of the intensity of the emitted beam on the detector, we performed simulations for the following situations: (a) a fully coherent source of Gaussian-distributed intensity, and (b) a fully incoherent source of Gaussian-distributed intensity. For both cases, the intensity distribution in the source plane $(\xi,\eta)$ is described by:

$$I_S(\xi,\eta) = I_{S,0} \exp\left(-\frac{\xi^2+\eta^2}{2\sigma_S^2}\right), \quad (2)$$

where $\sigma_S$ is the standard deviation, $(\xi,\eta)$ are the coordinates in the source plane and $I_{S,0}$ is the maximum of the intensity, $I_{S,0} = \frac{1}{2\pi\sigma_S^2}$ for a source with a total emission of 1. From Eq. (2) it follows that the amplitude of the wavefront distribution in the source plane is given by

$$A_S(\xi,\eta) = \sqrt{I_{S,0}} \exp\left(-\frac{\xi^2+\eta^2}{4\sigma_S^2}\right). \quad (3)$$

## 2.1. Coherent source

For a coherent source, each point on the source emits a wave in the same phase; for simplicity, we set the phase to 0. The wavefront distribution that propagates to the detector plane can be calculated by applying the Huygens-Fresnel principle:

$$U_D(X,Y) = -\frac{i}{\lambda}\iint U_S(\xi,\eta)\frac{\exp\left(ik\sqrt{(X-\xi)^2+(Y-\eta)^2+z^2}\right)}{\sqrt{(X-\xi)^2+(Y-\eta)^2+z^2}}d\xi d\eta, \quad (4)$$

where $z$ is the distance between the source and the detector. The profile of the intensity distribution in the detector plane for $\sigma_S = 2$ Å simulated according Eq. (4) is shown in Fig. 2, the curve labelled "Gaussian coherent source".

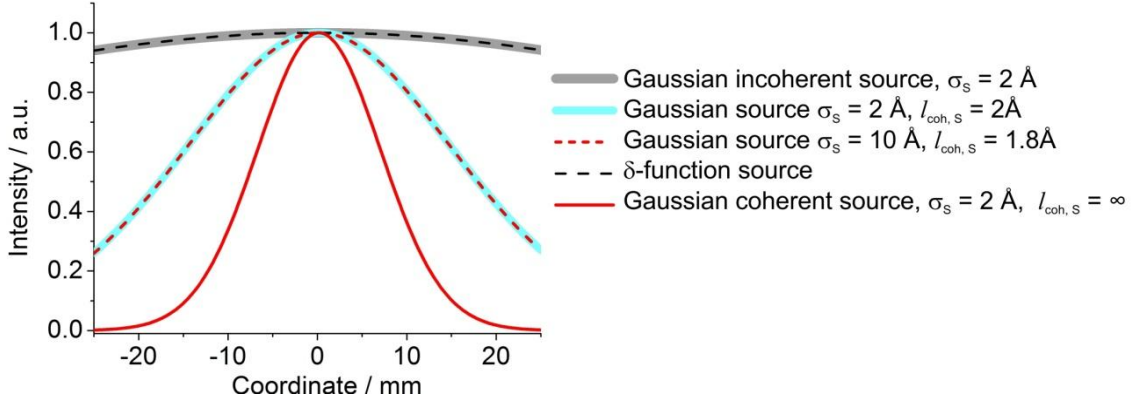

Fig. 2. Profiles of the intensity distributions in the detector plane for coherent, incoherent, partially coherent and $\delta$-function source. The kinetic energy of the electrons is 50 eV, and the source-to-detector distance is 0.1 m.

By substituting Eq. (3) into Eq. (4) and assuming that the detector plane can be considered as a far-field relative to the source plane, the following approximation can be applied $\sqrt{(X-\xi)^2+(Y-\eta)^2+z^2} \approx z+(X^2+Y^2)/(2z)-(X\xi+Y\eta)/z$, and we obtain the distribution of the electron wave in the detector plane:

$$U_D(X,Y) = -\frac{i4\pi\sigma_S^2}{\lambda z}\sqrt{I_{S,0}}\exp(ikz)\exp\left(\frac{i\pi}{\lambda z}(X^2+Y^2)\right)\exp\left(-\frac{4\pi^2\sigma_S^2(X^2+Y^2)}{(\lambda z)^2}\right), \quad (5)$$

where we employed the property of a Gaussian function that its Fourier transform is also a Gaussian function, as expressed by Eq. (A.1). From Eq. (5), we obtain the intensity distribution at the detector:

$$I_D(X,Y) = I_{D,0}\exp\left(-\frac{(X^2+Y^2)}{2\sigma_D^2}\right), \quad (6)$$

where we introduced the standard deviation of the intensity in the object plane

$$\sigma_D = \frac{\lambda z}{4\pi\sigma_S} \quad (7)$$

and $I_{D,0} = \left(\frac{4\pi\sigma_S^2}{\lambda z}\right)^2 I_{S,0}$. Thus, the source size $\sigma_S$ can be estimated directly from the background intensity distribution by fitting it with a two-dimensional Gaussian followed by applying Eq. (7). We hereafter refer to this approach as Method 2.

For example, Chang et al recorded a distribution of intensity from a single atom tip at a source-to-detector distance $z = 0.17$ m employing electrons of kinetic energy of 1400 eV, and fitted the background intensity distribution with a Gaussian function of standard deviation $\sigma_D = 4.5$ mm [7]. Assuming that the source was fully coherent, with these parameters according to Eq. (7), we obtain the source size $\sigma_S = 0.99$ Å, which is smaller than for example the atomic radius of tungsten (1.35 Å [13]) – the most common material for tip sources.

The divergence angle of the emission cone can be also estimated from fitting the background intensity distribution with a Gaussian distribution with variance $\sigma_D$. The emission angle is then evaluated as $\theta = \arctan(\sigma_D / z)$, as illustrated in Fig. 1. For example, for a single-atom tip described by Chang et al., we obtain $\theta = 1.5°$. Previously, Binh et al. experimentally studied emission from nanotips and reported the Gaussian-distributed intensity with the full width at the half maximum of 4° [12], which gives the variance of the Gaussian distribution $\theta = 1.7°$. This value agrees well with the emission angle reported by Chang et al [7].

## 2.2 Intensity distribution from an incoherent Gaussian-distributed intensity source

For an incoherent source, each point on the source emits a wave with a different phase. The intensity in the detector plane is then found as the sum of the intensities of the waves emitted from different points within the source:

$$I_D(X,Y) = \iint \left| -\frac{i}{\lambda} U_S(\xi,\eta) \cdot \frac{e^{ik\sqrt{(X-\xi)^2+(Y-\eta)^2+z^2}}}{\sqrt{(X-\xi)^2+(Y-\eta)^2+z^2}} \right|^2 d\xi d\eta. \quad (8)$$

The profile of the intensity distribution simulated in the detector plane for $\sigma_S = 2$ Å by employing Eq. (8) is shown in Fig. 2, the curve labelled "Gaussian incoherent source". Interestingly, the profile resembles the intensity distribution of a spherical wave. Thus, a wave emitted by a non-coherent source appears on a detector as if it is a spherical wave originating from a point-like source, despite the fact that its intensity distribution in the source plane is described by a Gaussian function.

By comparing the distributions obtained with coherent and incoherent sources, we conclude that if the intensity distribution in the detector plane is Gaussian-distributed, the source is coherent or partially coherent; such a source is not completely incoherent. Because the experimentally observed intensity distributions are mainly Gaussian-distributed, the physical sources are not completely incoherent, and are at least partially coherent. This property of "inner" coherence of emission tips was previously experimentally observed by Cho et al [6].

## 3. van Cittert-Zernike theorem
### 3.1. Statement of the theorem

The van Cittert-Zernike theorem considers an incoherent source and states that the mutual intensity function (MIF) in the far-field is given by the Fourier transform of the source intensity distribution. The theorem is also often re-phrased in terms of the complex coherence factor (CCF), which is a normalised MIF [2]:

$$\mu(\Delta x, \Delta y) = \frac{\exp(-i\psi) \int_{-\infty}^{+\infty}\int I(\xi,\eta)\exp\left(\frac{2\pi i}{\lambda z}(\xi\Delta x + \eta\Delta y)\right) d\xi d\eta}{\int_{-\infty}^{+\infty}\int I(\xi,\eta) d\xi d\eta}, \quad (9)$$

where

$$\psi = \frac{\pi}{\lambda z}\left[\left(x_2^2 + y_2^2\right) - \left(x_1^2 + y_1^2\right)\right] = \frac{\pi}{\lambda z}\left(\rho_2^2 - \rho_1^2\right), \quad (10)$$

$\rho_2$ and $\rho_2$ represent, respectively, the distances of the points $(x_2, y_2)$ and $(x_1, y_1)$ from the optical axis; $I(\xi,\eta)$ is the intensity distribution of the source; $(\xi,\eta)$ are the coordinates in the source plane; $(x_1, y_1; x_2, y_2)$ are the coordinates in the far-field plane; $\Delta x = x_2 - x_2$, $\Delta y = y_2 - y_1$; $\lambda$ is the wavelength and $z$ is distance from the source to the far-field plane. The factor $e^{-j\psi}$ can be neglected when the distance $z$ is so large that $z \gg \frac{2}{\lambda}\left(\rho_2^2 - \rho_1^2\right)$.

## 3.2. van Cittert-Zernike theorem applied to a uniform source

The van Cittert-Zernike theorem can be applied to a source of any intensity distribution. For example, for an incoherent source of uniform intensity of area $A_S$, by applying Eq. (9), we obtain the coherence area $A_C$ [2]:

$$A_C = \frac{(\lambda z)^2}{A_S}. \tag{11}$$

Assuming that both areas, the source area $A_S$ and the coherence area $A_C$, have a round shape, we obtain the radius of the incoherent source:

$$r_{\text{eff}} = \frac{\lambda z}{\pi R}, \tag{12}$$

where $R$ is the radius of the coherence area in the far-field, as illustrated in Fig. 1. The approach employing Eq. (12) corresponds to the aforementioned Method 1, which is often employed to estimate the effective source size of field emission tips [6], as for example in low-energy electron imaging [5, 7-8].

However, as already mentioned, the experimentally observed intensity distributions are mainly Gaussian-distributed, implying that the source is at least partially coherent. Therefore, it is not possible to interpret the experimentally observed intensity distributions as originating from a round-shaped incoherent source; thus, strictly speaking, the van Cittert-Zernike theorem cannot be applied.

## 3.3. van Cittert-Zernike theorem applied to a Gaussian-distributed intensity source

Here we consider an incoherent source with Gaussian-distributed intensity and derive its CCF, which will be important for further simulations at partial coherence. The intensity distribution of a Gaussian source is described by Eq. (2), where we assume a source with a total emission of 1, so that $I_{S,0} = \frac{1}{2\pi\sigma_S^2}$. Applying the van Cittert-Zernike theorem, the CCF distribution in the object plane ($x$, $y$) is obtained by calculating the Fourier transform of the intensity of the source:

$$\mu(\Delta x, \Delta y) = \int\int_{-\infty}^{+\infty} \frac{1}{2\pi\sigma_S^2} \exp\left(-\frac{\xi^2 + \eta^2}{2\sigma_S^2}\right) \exp\left(\frac{2\pi i}{\lambda z_0}(\xi\Delta x + \eta\Delta y)\right) d\xi d\eta =$$
$$= \exp\left(-2\pi^2\sigma_S^2 \frac{(\Delta x^2 + \Delta y^2)}{(\lambda z_0)^2}\right) = \exp\left(-\frac{\Delta x^2 + \Delta y^2}{2l_{\text{coh}}^2}\right) \tag{13}$$

where we have substituted Eq. (2) into Eq. (9) and introduced the coherence length:

$$l_{\text{coh}} = \frac{\lambda z_0}{2\pi\sigma_S} \quad (14)$$

where $z_0$ is the distance between the source plane and the object plane. The expression for the CCF given by Eq. (13) is an exact far-field solution according to the van Cittert-Zernike theorem [2, 4] for an incoherent source with a Gaussian distribution of intensity. The CCF as expressed by Eq. (13) implies that the coherence between the waves originating from different points of the source does not abruptly disappear when the distance between those points exceeds some certain threshold distance. In contrast, there will always be interference, but the contrast (visibility) of the interference pattern will decrease as the distance between the points increases. $l_{\text{coh}}$ defines the distance at which the interference contrast will drop from 1 to 0.6. For example, in point-projection microscopy or inline holography [14], where an object is placed into a divergent electron beam, to ensure coherent imaging, the coherence length in the object plane $l_{\text{coh}}$ must not only be comparable to the object size (to ensure that all waves scattered by the object will coherently interfere); but also $l_{\text{coh}}$ must be several times larger than the object size to ensure that there is also interference with the non-scattered, reference wave.

## 4. Partially coherent Gaussian-distributed intensity source
### 4.1 Simulation as convolution

The intensity distribution of a partially coherent wave in the far-field can be represented as the convolution between the intensity distribution of a fully coherent wave in the far-field with the Fourier transform of the complex coherence factor (CCF) function $\mu(\Delta x, \Delta y)$, which we denote as $M(K_x, K_y)$ [15-16]:

$$I(K_x, K_y) = I_{\text{coh}}(K_x, K_y) \otimes M(K_x, K_y), \quad (15)$$

where $\otimes$ denotes convolution and the $K$-coordinates are the unit-less emission vector coordinates defined as follows:

$$\vec{K} = \left(\frac{X}{R}, \frac{Y}{R}, \frac{z}{R}\right), \quad R = X^2 + Y^2 + z^2. \quad (16)$$

The resulting intensity distribution in $(K_x, K_y)$-coordinates can be re-calculated in $(X, Y)$-coordinates by applying corresponding coordinate transformation [17]. Equation (15) gives the resulting distribution, which corresponds to the situation where an intensity distribution obtained from a fully coherent source is altered by taking into account that the waves emitted from different points within the source interfere at a degree of coherence that decreases as the distance between the points increases.

A profile of simulated intensity originating from a Gaussian-distributed intensity source of $\sigma_S = 2$ Å with partial coherence in the source plane $l_{\text{coh, S}} = 2$ Å is shown in Fig. 2. It lies between the profiles from a fully coherent source with $\sigma_S = 2$ Å and an incoherent source.

## 4.2. Analytical solution

The intensity distribution of a partially coherent source can be obtained directly by calculating the convolution in Eq. (15) analytically. For a fully coherent source, the intensity distribution $(X,Y)$ in the detector plane is given by Eq. (1) and its Fourier transform can be calculated analytically:

$$\iint I_D(X,Y) \exp\left(-\frac{2\pi i}{\lambda z}(xX + yY)\right) dXdY = 2\pi\sigma_D^2 \cdot I_{D,0} \exp\left(-2\pi^2\sigma_D^2 \frac{x^2 + y^2}{(\lambda z)^2}\right),$$

the Fourier transform of $M(K_x, K_y)$ is the CCF function given by:

$$\mu(x,y) = \exp\left(-\frac{x^2 + y^2}{2l_{\text{coh}}^2}\right),$$

and their product is given by

$$2\pi\sigma_D^2 \cdot I_{D,0} \exp\left(-2\pi^2\sigma_D^2 \frac{x^2 + y^2}{(\lambda z)^2}\right) \exp\left(-\frac{x^2 + y^2}{2l_{\text{coh}}^2}\right) = 2\pi\sigma_D^2 \cdot I_{D,0} \exp\left(-\frac{x^2 + y^2}{2\chi^2}\right),$$

where we introduced

$$2\pi^2\sigma_D^2 \frac{1}{(\lambda z)^2} + \frac{1}{2l_{\text{coh}}^2} = \frac{1}{2\chi^2}.$$

The inverse Fourier transform gives:

$$I_D^{(\text{Partial})}(X,Y) = \frac{1}{(\lambda z)^2} 2\pi\sigma_D^2 \cdot I_{D,0} \iint \exp\left(-\frac{x^2 + y^2}{2\chi^2}\right) \exp\left(\frac{2\pi i}{\lambda z}(xX + yY)\right) dxdy =$$

$$= \frac{4\pi^2\sigma_D^2\chi^2}{(\lambda z)^2} I_{D,0} \exp\left(-2\pi^2\chi^2 \frac{X^2 + Y^2}{(\lambda z)^2}\right) \sim$$

$$\sim I_{D,0} \exp\left(-\frac{X^2 + Y^2}{2\sigma_{\text{exp}}^2}\right),$$

(17)

where $\sigma_{exp}$ is standard deviation of the experimentally observed Gaussian distribution:

$$\sigma_{exp}^2 = \frac{(\lambda z)^2}{4\pi^2 \chi^2} = \left(\frac{\lambda z}{2\pi}\right)^2 \left(\frac{1}{(2\sigma_S)^2} + \frac{1}{l_{coh}^2}\right), \quad (18)$$

and where we used Eq. (7). Thus, the intensity distribution on the detector is described by a two-dimensional Gaussian function whose standard deviation $\sigma_{exp}$ is defined by both the source size $\sigma_S$ and the coherence length in the source plane $l_{coh,S}$.

From Eq. (18), the following practical formulae for estimations of $\sigma_S$ and $l_{coh,S}$ from a known $\sigma_{exp}$ can be derived:

$$\sigma_S \leq \frac{\lambda z}{4\pi\sigma_{exp}} \quad \text{and} \quad l_{coh,S} \leq \frac{\lambda z}{2\pi\sigma_{exp}}. \quad (19)$$

The following conclusions can be made from Eq. (18). (1) For a fully coherent source ($l_{coh,S} = \infty$) the second term in Eq. (18) turns to zero and the source size $\sigma_S$ can be directly evaluated from $\sigma_{exp}$; (2) When the coherence length $l_{coh,S}$ approaches zero, $\sigma_{exp}$ becomes mainly defined by the second term and approaches infinity $\sigma_{exp} \to \infty$, and the intensity distribution resembles that of a spherical wave: almost a constant intensity, independent of the source size.

From Eq. (18) it follows that from the intensity distribution alone, it is not possible to separate the two terms nor to evaluate the source size or the coherence length. For example, the two following sets of parameters: $\sigma_S = 2$ Å, $l_{coh,S} = 2$ Å, and $\sigma_S = 10$ Å, $l_{coh,S} = 1.8$ Å, according to Eq. (18), lead to the same $\sigma_{exp}$. The intensity distributions with these parameters were simulated and their profiles shown in Fig. 2 are identical.

## 5. Evaluation of effective source size from biprism effect

Conventionally, the effective source size is evaluated from the extent of an experimental interference pattern formed by electron beam diffraction on a charged wire, or from the so-called biprism effect [18]. For low-energy electrons, typically less than 1 keV, a nanotube is placed in front of a tip that acts as a charged wire, because it creates a potential distribution that bends the trajectories of the emitted electrons [19-20]. In this section, we present the simulated biprism interference pattern, evaluate the effective source size, and compare it with the pre-defined value.

## 5.1. Coherent source

The biprism interference pattern, or hologram, is simulated as:

$$H_{\text{coh}}(K_x, K_y) = \left| \frac{1}{\lambda z} \iint U_O(x,y) t(x,y) e^{-ik(K_x x + K_y y)} dx dy \right|^2 \quad (20)$$

where $U_O(x,y)$ is the illuminating wavefront and $t(x,y)$ is the transmission function in the object domain. The incident wavefront in the object plane $U_O(x,y)$ is calculated by applying the Huygens-Fresnel principle:

$$U_O(x,y) = -\frac{i}{\lambda z_0} \iint U_S(\xi, \eta) \cdot e^{ik\sqrt{(x-\xi)^2 + (y-\eta)^2 + z_0^2}} d\xi d\eta. \quad (21)$$

By substituting $U_S(\xi, \eta)$ given by Eq. (3) into Eq. (21), and assuming that the object plane can be considered as a far-field relative to the source plane, so that $\sqrt{(x-\xi)^2 + (y-\eta)^2 + z_0^2} \approx z_0 + (x^2 + y^2)/(2z_0) - (x\xi + y\eta)/z_0$, we obtain the distribution of the electron wave in the object plane:

$$\begin{aligned}
U_O(x,y) &= -\frac{i}{\lambda z} \exp\left(\frac{2\pi i}{\lambda} z_0\right) \exp\left[\frac{\pi i}{\lambda z_0}(x^2 + y^2)\right] \\
&\times \iint U_S(\xi, \eta) \exp\left[-\frac{2\pi i}{\lambda z_0}(x\xi + y\eta)\right] d\xi d\eta \\
&= -\frac{i}{\lambda z} \sqrt{I_{S,0}} \exp\left(\frac{2\pi i}{\lambda} z_0\right) \exp\left[\frac{\pi i}{\lambda z_0}(x^2 + y^2)\right] \\
&\times \iint \exp\left[-\frac{\xi^2 + \eta^2}{4\sigma_S^2}\right] \exp\left[-\frac{2\pi i}{\lambda z_0}(x\xi + y\eta)\right] d\xi d\eta \\
&= -\frac{i4\pi\sigma_S^2}{\lambda z} \sqrt{I_{S,0}} \exp\left(\frac{2\pi i}{\lambda} z_0\right) \exp\left[\frac{\pi i}{\lambda z_0}(x^2 + y^2)\right] \exp\left[-\frac{4\pi^2 \sigma_S^2 (x^2 + y^2)}{(\lambda z_0)^2}\right] \sim \\
&\sim \exp\left[-\frac{(x^2 + y^2)}{2\sigma_O^2}\right],
\end{aligned} \quad (22)$$

where we employed the result that the Fourier transform of a Gaussian function is also a Gaussian function, as expressed by Eq. (A.1), and introduced $\sigma_O$:

$$\sigma_O = \frac{\lambda z_0}{2\pi \sigma_S}. \quad (23)$$

The transmission function $t(x,y)$ in Eq. (20) has an amplitude of 1 everywhere except for the region of the wire, where the amplitude is set to 0, mimicking an opaque wire. The phase distribution is described by a prism-like distribution that is zero within the wire and decays outwards from the wire, as discussed in [20]. The simulated hologram is transformed from $(K_x, K_y)$ to $(X, Y)$ coordinates [17], giving $H_{\text{coh}}(X,Y)$. In the simulations, it is assumed that the wire is 1.7 nm in diameter, has a linear charge of $Q = 0.15$ $e$/nm, and is placed at a distance of $z_0 = 200$ nm from the source. The other parameters of simulations are: the kinetic energy of the electrons is 50 eV (wavelength $\approx$ 1.73 Å) and the source-to-detector distance is $z = 0.1$ m. The experimental data always contain some additional constant background that is not included in the simulation. The holograms shown here rather indicate the extent of the interference pattern; for example, the hologram simulated at $\sigma_S = 3$ Å and shown in Fig. 3(a) exhibits 3–5 fringes observed with the naked eye.

Figure 3 shows simulated holograms at different source sizes of a coherent source: $\sigma_S = 1, 2, 3$ Å and $\delta$-function. From the results shown in Fig. 3(a)–(c), it is evident that, in the case of a finite-sized source, the spread of the biprism interference pattern is mainly limited by the modulating Gaussian distribution of the reference wave, with the distribution variance $\sigma_D$ inversely proportional to $\sigma_S$, as described by Eq. (7). The maxima in the interference patterns are observed at the very edge of the intensity distribution for all source sizes. Employing Method 1, we estimate the source size $r_{\text{eff}} \approx$ 0.96, 3.56, 4.41 and 5.16 Å for the source size $\sigma_S = \delta$–function, 1, 2 and 3 Å, respectively. These values are about double the initial $\sigma_S$ values.

More precise estimations of the source size can be obtained with Method 2. The low-pass filtered images exhibit a Gaussian-distributed intensity that approximately corresponds to the reference beam intensity. The intensity profiles of the low-pass filtered biprism interference patterns are shown with dashed lines in Fig. 3(b). With Method 2, we obtain $\sigma_S \approx$ 1.0, 1.9 and 2.5 Å for the source size $\sigma_S$ = 1, 2 and 3 Å, respectively. These values fit the initial source size values well.

The holograms shown in Fig. 3(a)–(c) exhibit one important property: their contrast is less than 1. The holograms' profiles resemble those of the experimental holograms of nanotubes presented and discussed in previous works [7-8]. Chang et al. discussed the reason why the visibility, or contrast, in experimental holograms does not reach 1, explaining it in terms of the energy spread of the electron source by inelastic scattering by the single-walled nanotube bundle [7]. However, as can be seen from Fig. 3(a)–(c), the visibility of the simulated hologram is also less than 1, though a monochromatic and noise-free electron beam and only elastic scattering were assumed in the simulation. Thus, the intensity distributions shown in Fig. 3(a)–(c) agree well with the previously reported experimental results [7], and their visibility < 1 can be explained solely by the Gaussian intensity distribution of the source.

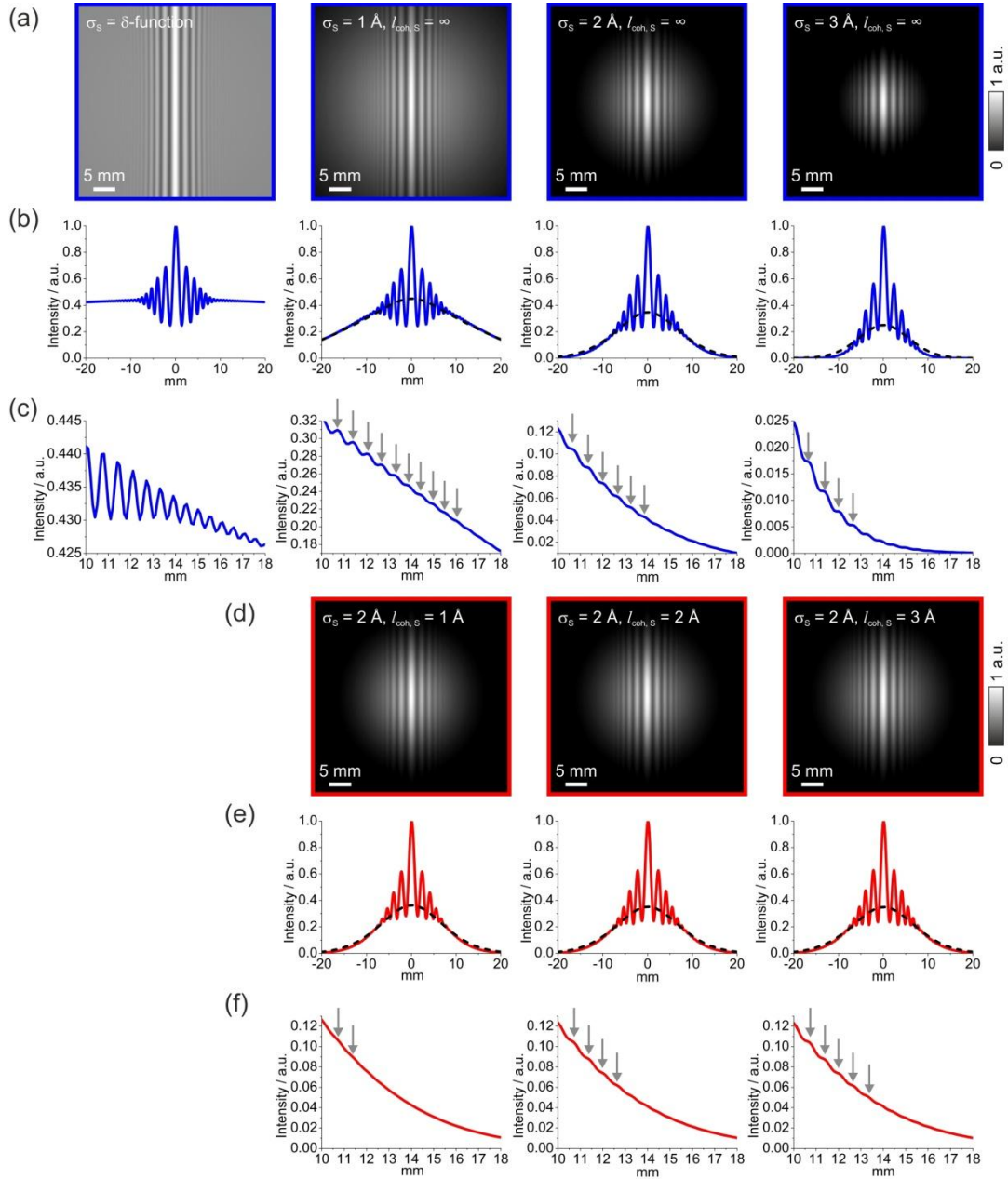

Fig. 3. Simulated holograms of a charged wire. (a)–(c): Simulations for a coherent source at different source sizes $\sigma_S = \delta$–function, 1 Å, 2 Å and 3 Å. (a): Simulated holograms, each normalised to its maximum value. (b): Profiles through the centre of the holograms shown in (a). (c): Magnified regions shown in (b) at the edge of the profiles. (d)–(f): Simulations for a partially coherent source of $\sigma = 2$ Å at different coherent length $l_{\text{coh,S}} = 1$, 2 and 3 Å. (d): Simulated holograms, each normalised to its maximum value. (e): Profiles through the centre of the holograms shown in (d). (f): Magnified regions shown in (e) at the edge of the profiles. The arrows in (c) and (f) indicate the maxima in the interference pattern. The parameters of the simulations are: the energy of the electrons is 50 eV, the source-to-sample distance is 200 nm, and the source-to-detector distance is 0.1 m.

### 5.2. Partially coherent source

The holograms acquired with a partially coherent source can be simulated as a convolution, similar to Eq. (15):

$$H(K_x, K_y) = H_{\text{coh}}(K_x, K_y) \otimes M(K_x, K_y). \tag{24}$$

Here, the function $M(K_x, K_y)$ is the Fourier transform of the CCF function that is given by Eq. (18), and where the coherence length $l_{\text{coh,O}}$ is related to the object plane. According to Pozzi [21], the ratio between the beam size and the coherence length is constant along the beam propagation. The beam size in the object plane is determined by the standard deviation, as expressed by Eq. (23), $\sigma_O = \dfrac{\lambda z_0}{2\pi\sigma_S}$. This gives the coherence length in the object plane:

$$l_{\text{coh,O}} = \frac{\lambda z_0}{2\pi\sigma_S^2} l_{\text{coh,S}}. \tag{25}$$

Figure 3(d)–(f) shows holograms simulated by employing Eq. (24), with the source size $\sigma_S = 2$ Å, and the coherence length is the source plane: $l_{\text{coh,S}} = 1$, 2 and 3 Å, which in the object plane at $z_0 = 200$ nm from the source plane turns into a coherence length of $l_{\text{coh,O}} =$ 13.8, 27.6 and 41.4 nm, respectively. From Figure 3 (d)–(f) it is apparent that when the coherence length $l_{\text{coh,S}}$ is less than the source size $\sigma_S$, the interference pattern loses its contrast, and fewer fringes are observed on the detector. The source sizes estimated by Method 1 are larger than the pre-defined values: $r_{\text{eff}} \approx$ 6.49 Å, 5.25 Å, and 4.41 Å for the coherence lengths of $l_{\text{coh,S}} = 1$ Å, 2 Å and 3 Å, respectively. Again, more precise estimations of the source size can be obtained with Method 2. The intensity profiles of the low-pass filtered biprism interference patterns are shown with dashed lines in Fig. 3(e). With Method 2,

we obtain $\sigma_S \approx 1.9$, 1.9 and 1.9 Å for the source size $\sigma_S = 2$ Å and the coherence length $l_{\text{coh},S} = 1$ Å, 2 Å and 3 Å, respectively, which match the predefined values well.

From the results presented in Fig. 3, it is evident that the main factor limiting the extent of the interference pattern on the screen is the fact that the electron beam intensity is Gaussian-distributed. The effect of finite coherence length is a less limiting effect, as can be seen from comparing the biprism interference pattern simulated for $\sigma_S = 2$ Å at infinite and finite coherence lengths, shown in Fig. 3.

## 6. Effect of source size on resolution

An elegant way of realising imaging with a coherent electron beam is via in-line Gabor holography [22-23] (also called point-projection imaging [24]), which does not require any lenses, and thus is free of aberrations. With the invention of very sharp and bright tips [9], in-line holograms with rich interference patterns became available [14, 25-26], and objects reconstructed at the nanometre resolution from their holograms were reported [20, 27-40]. Here, estimation of the effective source size is mainly essential for the evaluation of the resolution of the imaged objects. The divergence angle of the emitted electron beam as a factor limiting the extent of the hologram was mentioned by Stevens [41]. We now investigate the resolution of objects reconstructed from holograms simulated at different source sizes.

We consider a fully coherent source of Gaussian-distributed intensity. As an object for testing the resolution, we consider an opaque object in the form of three sets of bars of equal width and separation: $w_1 = 2$ Å, $w_2 = 4$ Å and $w_3 = 6$ Å, as shown in Fig. 4(a). We simulate holograms with the sources of three different sizes: $\sigma_{S,1} = 1$ Å, $\sigma_{S,2} = 2$ Å and $\sigma_{S,3} = 3$ Å. The other parameters of the simulation are: the energy of the electrons is 50 eV, the source-to-sample distance is 230 nm, and the source-to-detector distance is 0.1 m. The holograms $H_{\text{coh}}(K_x, K_y)$ are simulated as the analytical solution of diffraction on a rectangular object. The simulated hologram sare shown in Fig. 4(b).

The numerical reconstruction of digital holograms consists of multiplication of the hologram with the reference wave $e^{ikR}/R$, where $R$ is defined in Eq. (16), followed by back propagation to the object plane described by the Huygens-Fresnel principle [42-43]:

$$U(x, y, z_0) = \frac{i}{\lambda} \iint H_{\text{coh}}(X, Y) \frac{e^{-ik|\vec{R}-\vec{r}|}}{|\vec{R}-\vec{r}|} \frac{e^{ikR}}{R} dX dY, \qquad (26)$$

where $U(x, y, z_0)$ is the reconstructed wavefront in the object plane and $\vec{r} = (x, y, z_0)$.

Applying approximation $|\vec{R} - \vec{r}| \approx R - \vec{R}\vec{r}/R = R - \vec{K}\vec{r}$, where $\vec{K}$ is defined in Eq. (16), we rewrite Eq. (26) as:

$$U(x, y, z_0) \approx \frac{i}{\lambda} \iint H_{\text{coh}}(K_x, K_y) \cdot e^{ik\vec{K}\vec{r}} dK_x dK_y, \qquad (27)$$

and when expanded:

$$U(x, y, z_0) \approx \frac{i}{\lambda} \iint H_{\text{coh}}(K_x, K_y) \cdot e^{\frac{2\pi i}{\lambda} z_0 \sqrt{1-K_x^2-K_y^2}} e^{\frac{2\pi i}{\lambda}(K_x x + K_y y)} dK_x dK_y. \qquad (28)$$

The transmission function in the object plane is then obtained by division with the incident wave $e^{ikr}/r$:

$$t(x, y, z_0) = U(x, y, z_0) r e^{-ikr}, \qquad (29)$$

where $r = \sqrt{x^2 + y^2 + z_0^2}$. The reconstructed transmission functions are shown in Fig. 4(c). The reconstructions shown in Fig. 4(c) demonstrate that the resolution becomes worse with increasing source size.

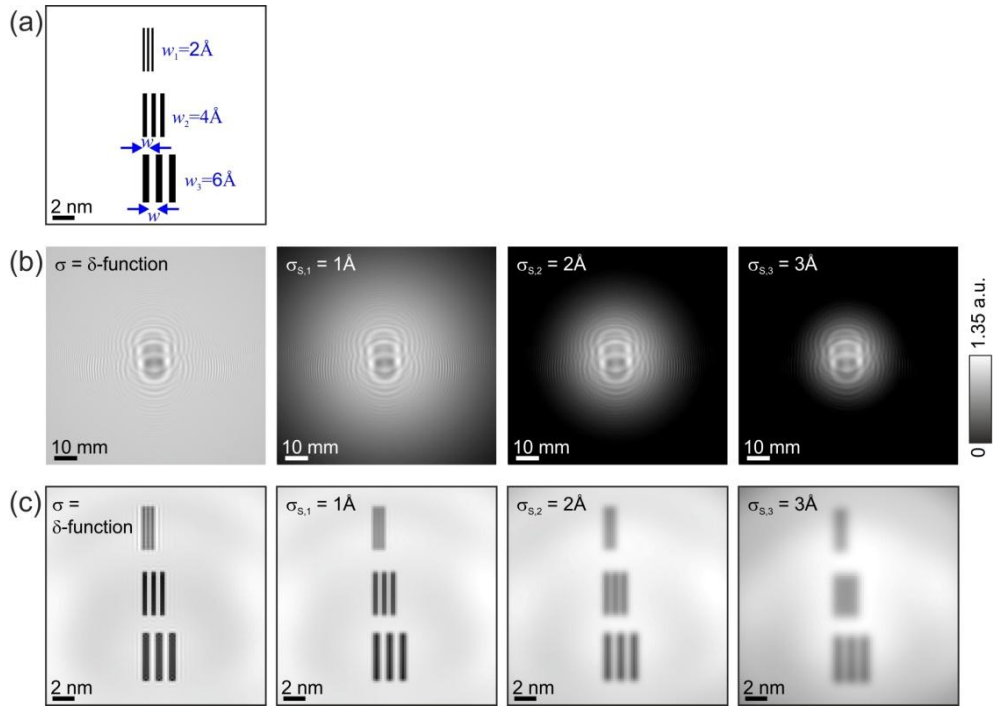

Fig. 4. Effect of source size on the resolution of objects recorded from their inline holograms. (a) Test object consisting of bars of equal width and separation: $w_1 = 2$ Å, $w_2 = 4$ Å and $w_3 = 6$ Å. (b) Corresponding simulated holograms $H_{\text{coh}}(X, Y)$,

where the source was assumed to be described by a $\delta$-function and by a two-dimensional Gaussian function with standard deviation $\sigma_{S,1} = 1$ Å, $\sigma_{S,2} = 2$ Å and $\sigma_{S,3} = 3$ Å. The parameters of the simulations are: the energy of the electrons is 50 eV, the source-to-sample distance is 230 nm, and the source-to-detector distance is 0.1 m. (c) Objects reconstructed from the holograms depicted in (b), amplitude distributions are shown.

The intrinsic resolution of an inline hologram is given by [40, 44-45]:

$$R_{\text{intrinsic}} = \frac{\lambda z}{S_H}, \tag{30}$$

where $z$ is the distance between the source and the detector and $S_H$ is the hologram size. The intrinsic resolution calculated according to Eq. (30) is $R_{\text{intrinsic}} = 1.8$ Å for the parameters listed above (the energy of the electrons is 50 eV, the source-to-sample distance is 230 nm, the source-to-detector distance is $z = 0.1$ m and the imaged sample area $200 \times 200$ nm$^2$ gives the hologram size $S_H = 96$ mm. From the reconstructions shown in Fig. 4(c), it is apparent that the finest bars $w_1 = 2$ Å are only resolved in the case of an ideal $\delta$-function source, which corresponds to the limit of the achievable resolution $R_{\text{intrinsic}} = 1.8$ Å. All other holograms simulated for realistic sources sizes exhibit worse resolution. From the reconstructions shown in Fig. 4(c), it is apparent the source with $\sigma_1 = 1$ Å, which corresponds to a single atom source, can produce a hologram where the smallest objects of $w_2 = 4$ Å separation can be resolved. A source with $\sigma_2 = 2$ Å can produce a hologram where the smallest objects of $w_3 = 6$ Å separation can be resolved. A source with $\sigma_3 = 3$ Å can produce a hologram where objects of even $w_3 = 6$ Å separation are barely resolved.

The conventional formula for resolution, Eq. (30), does not provide the correct estimation in the case of a finite-sized source. From the simulated holograms shown in Figs. (3) and (4), it is evident that in the case of a finite-sized source, the spread of the hologram is limited by the Gaussian distribution; thus, the radius of the effective detector area is given by:

$$R_{\text{eff}} = \frac{2\pi \sigma_D}{\omega} = \frac{\lambda z}{2\omega \sigma_S} \tag{31}$$

where we used Eq. (7). Here, $\omega$ is a coefficient that helps to select the cut-off coordinate limited by the Gaussian distribution. The Gaussian distribution is broad; therefore, the cut-off coordinate is not well pronounced, but it can be selected at a point where the amplitude of the Gaussian drops to a certain value. For example, when $\omega = 1$, $R_{\text{eff}}$ limits the area where the Gaussian amplitude drops to 0.6 of its maximum value.

By setting $S_H = 2R_{\text{eff}}$ and substituting Eq. (31) into Eq. (30), we obtain:

$$R_{\text{intrinsic}} = \frac{\lambda z}{2R_{\text{eff}}} = \omega \sigma_S. \tag{32}$$

Equation (32) confirms a well-known fact that the resolution is proportional to the source size. When $\omega = 2$, $R_{\text{intrinsic}} = 2\sigma_S$, which well agrees with the resolution estimated from the simulated holograms shown in Fig. 4 (2, 4 and 6 Å for $\sigma_3 = 1$, 2 and 3 Å, respectively).

## 7. Effect of vibrations

Until now we have considered the situation where the source and the sample are spatially fixed, which results in a fixed interference pattern on the detector. This is, however, not a situation in a realistic experiment, since any source or sample is subject to mechanical vibrations. The question is: which amplitude of vibrations leads to which loss of resolution? In this section, we quantitatively study the degradation of resolution as a function of the amplitude of mechanical vibrations. A lateral shift of either the source or the sample by $\Delta r$ results in a shift of the entire interference pattern by

$$\Delta R = M \cdot \Delta r \tag{33}$$

where $M$ is the magnification factor, equal to the ratio of the source-to-detector distance to the source-to-sample distance.

Should the source or the sample constantly vibrate, the recorded hologram will be a superposition of all shifted holograms; thus the higher-order fringes will be smeared out, giving a blurred hologram. The simulated results of such vibrations are demonstrated in Fig. 5, where the source size was selected as for a single atom tip $\sigma_S = 1$ Å and a situation when the source was vibrating within $\Delta r = \pm 1$ nm is studied. In the left of Fig. 5(a), the original hologram is shown: there are about 20 interference fringes (counted from the centre). In the right Fig. 5(a), the blurred hologram is shown: there are only up to seven interference fringes. Thus, even when the source provides a fully coherent beam, unavoidable vibrations smear out higher-order interference fringes, and therefore degrade the resolution.

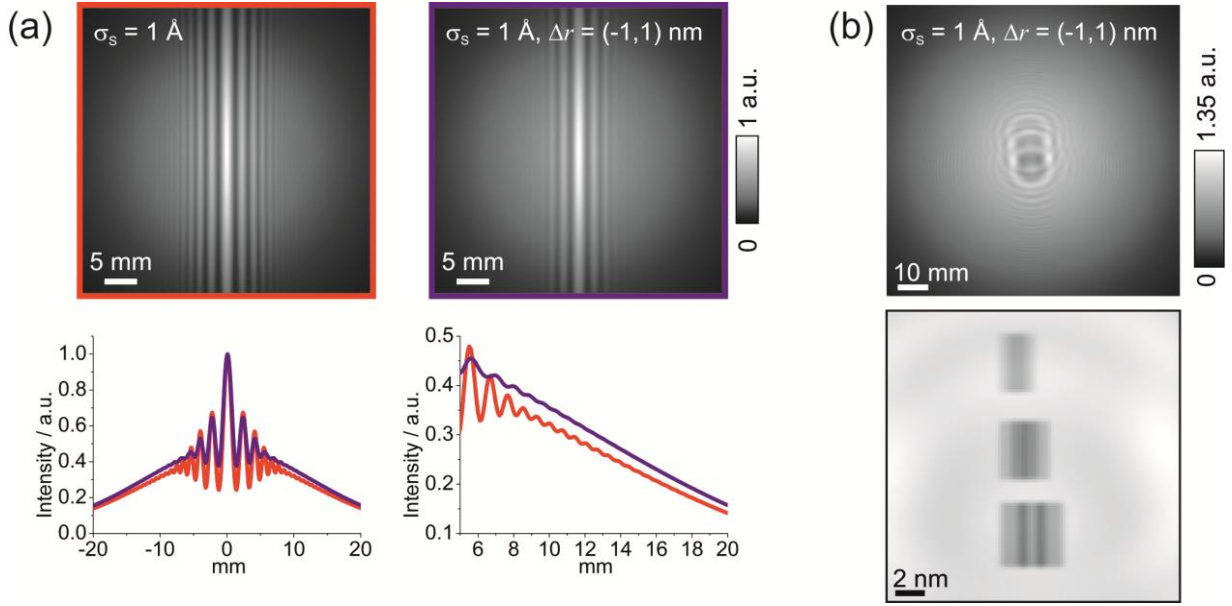

Fig. 5. Effect of mechanical vibrations. (a) Top: Hologram simulated for a fixed source and a source, vibrating within $\Delta r = \pm 1$ nm. Bottom: related intensity profiles though the centre of the holograms and their magnified regions at the edge of the interference pattern. The parameters of the simulations are: the energy of the electrons is 50 eV, the source-to-sample distance is 200 nm, and the source-to-detector distance is 0.1 m. (b) Top: simulated hologram of test objects bar as in Fig. 4, with the source size $\sigma_S = 1$ Å, and vibrating within $\Delta r = \pm 1$ nm. Bottom: object reconstructed from the hologram. The parameters of the simulations are: the energy of the electrons is 50 eV, the source-to-sample distance is 230 nm, and the source-to-detector distance is 0.1 m.

By applying Method 1, we obtain the effective source sizes for the holograms shown in Fig. 5(a): $r_{eff} \approx 3.6$ and 7.0 Å, respectively. Method 2 gives $\sigma_S = 1.05$ Å for both holograms, which is the correct pre-defined value. However, obviously, one should expect a different resolution to the original and blurred holograms. To study the effect of the vibration on the resolution, we consider the test bar object. The hologram simulated for a fixed $\sigma_S = 1$ Å source was shown previously in Fig. 4. The smallest objects that can be resolved in its reconstruction are of $w_2 = 4$ Å separation. The hologram simulated for a source vibrating within $\Delta r = \pm 1$ nm and its reconstruction are shown in Fig. 5(b). The reconstruction does not show any correctly reconstructed bars, and thus its resolution is worse than 6 Å. By taking into account that $r_{eff}$ estimated by Method 1 is typically about 2–3 times larger than $\sigma_S$, we can re-write the formula for the resolution given by Eq. (32) as

$$R_{\text{intrinsic}} \approx r_{\text{eff}}. \tag{34}$$

Equation (34) is a practical formula for estimating the intrinsic resolution of an experimental imaging system, where $r_{\text{eff}}$ is evaluated from the extent of the interference pattern by Method 1. Equation (34) gives $R_{\text{intrinsic}} \approx 3.56$ and 6.98 Å, for fixed and vibrating sources, respectively; these values match the quality of the corresponding reconstructions (cf. results shown in Figs. 4 and 5).

## 8. Conclusions

In conclusion, our simulations show that the Gaussian-distributed intensity of an electron wave emitted from a nano- or sub-nano-sized source observed on the detector implies that the source is at least partially coherent. This confirms previous experimental observations [6]. The fact that the source is coherent, however, complicates the application of the van Cittert-Zernike theorem, which assumes an incoherent source. We have proposed another approach of estimating the effective source size: by assuming that the source is fully coherent, and that its intensity is Gaussian-distributed. The effective source size is then evaluated from the distribution of the intensity on the detector. This approach is justified for nano-tips where the coherence length usually exceeds its lateral dimensions, and where the emitted wave can be considered as fully coherent. The values of the effective source size obtained by this approach agree with the pre-defined values well. We also evaluated the effective source size by the conventional approach, from the extent of the interference pattern, and by applying the van Cittert Zernike theorem. The obtained values are 2–3 times larger than the pre-defined values; however, they serve as a good measure of the intrinsic resolution of the imaging system.

## Appendix I. Fourier transform of a Gaussian function

The Fourier transform of a Gaussian function is also a Gaussian function:

$$\iint \exp\left(-\frac{x^2 + y^2}{2\sigma^2}\right) \exp\left(-2i\pi(x\alpha + y\beta)\right) \mathrm{d}x \mathrm{d}y = 2\pi\sigma^2 \exp\left(-2\pi^2\sigma^2(\alpha^2 + \beta^2)\right). \tag{A.1}$$